\documentclass[aps,twocolumn,prb,showpacs,floatfix,amsmath,amssymb,superscriptaddress]{revtex4-1}
\usepackage{graphicx}
\usepackage{bm}
\usepackage{hyperref}
\usepackage{color}
\usepackage{natbib}
\usepackage{changes,cancel,amsmath, amssymb}
\usepackage{braket,float,soul}
\usepackage{comment}
\usepackage{ulem}

\definecolor{ao(english)}{rgb}{0.0, 0.5, 0.0}

\DeclareMathOperator*{\SumInt}{%
\mathchoice%
  {\ooalign{$\displaystyle\sum$\cr\hidewidth$\displaystyle\int$\hidewidth\cr}}
  {\ooalign{\raisebox{.14\height}{\scalebox{.7}{$\textstyle\sum$}}\cr\hidewidth$\textstyle\int$\hidewidth\cr}}
  {\ooalign{\raisebox{.2\height}{\scalebox{.6}{$\scriptstyle\sum$}}\cr$\scriptstyle\int$\cr}}
  {\ooalign{\raisebox{.2\height}{\scalebox{.6}{$\scriptstyle\sum$}}\cr$\scriptstyle\int$\cr}}
}

\graphicspath{{./Figures-bs/} } 
\begin{document}
%%%%%%%%%%%%%%%%%%%%%%
%%%% short version of the title %%%%%%
%%%%%%%%%%%%%%%%%%%%%%
%\title{Negative absorption   induced by a local quench \\ as a signature  of Generalized Gibbs Ensemble  }
%%%%%%%%%%%%%%%%%%%%%%
%%%% long version 1 of the title %%%%% 
%%%%%%%%%%%%%%%%%%%%%%
\title{Signature  of  Generalized Gibbs Ensemble  Deviation from Equilibrium: \\ Negative Absorption Induced by a Local Quench}
%%%%%%%%%%%%%%%%%%%%%%
%%%% long version 2 of the title %%%%% 
%%%%%%%%%%%%%%%%%%%%%%
%\title{Signature  of minimal Generalized Gibbs Ensemble  deviation from Equilibrium:\\ Optical gain   induced by a local quench}

%%%%%%%%%%%%%%%%%%%%%%%%%%%%%
\author{{Lorenzo  Rossi}}
\email{lorenzo.rossi@polito.it}
\affiliation{Dipartimento di Scienza Applicata e Tecnologia, Politecnico di Torino, 10129 Torino, Italy}

\author{{Fabrizio Dolcini}}
\affiliation{Dipartimento di Scienza Applicata e Tecnologia, Politecnico di Torino, 10129 Torino, Italy}

\author{{Fabio Cavaliere}}
\affiliation{Dipartimento di Fisica, Universit\`a di Genova, 16146 Genova, Italy}
\affiliation{SPIN-CNR, 16146 Genova, Italy}
 
\author{Niccol\`o~Traverso~Ziani}
\affiliation{Dipartimento di Fisica, Universit\`a di Genova, 16146 Genova, Italy}
\affiliation{SPIN-CNR, 16146 Genova, Italy}

\author{{Maura Sassetti}}
\affiliation{Dipartimento di Fisica, Universit\`a di Genova, 16146 Genova, Italy}
\affiliation{SPIN-CNR, 16146 Genova, Italy}

\author{{Fausto Rossi}}
\affiliation{Dipartimento di Scienza Applicata e Tecnologia, Politecnico di Torino, 10129 Torino, Italy}

\begin{abstract}
When a parameter quench is performed in an isolated quantum system with a complete set of constants of motion, its out of equilibrium dynamics  is considered to be well captured by the Generalized Gibbs Ensemble (GGE), characterized by a set $\{\lambda_\alpha\}$ of coefficients  related to the constants of motion. We determine the most elementary  GGE  deviation  from the equilibrium distribution that leads to detectable effects. By quenching a suitable local  attractive potential in a one-dimensional electron system, the resulting GGE  differs from   equilibrium  by only one single $\lambda_{\alpha}$, corresponding to the emergence of an only partially occupied bound state lying below a fully occupied continuum of states. The effect is shown to  induce  optical gain, i.e., a negative peak in the absorption spectrum, indicating the stimulated emission of   radiation, enabling one to identify GGE signatures in fermionic systems through optical measurements. We discuss the implementation in realistic setups.
\end{abstract}

\maketitle
%%%%%%%%%%%%%%%%%%%%%%%%%%%%%%%%%%%% 
%%%%%%%%%%%%%%%%%%%%%%%%%%%%%%%%%%%% 
%%%%%%%%      INTRODUCTION         %%%%%%%%%% 
%%%%%%%%%%%%%%%%%%%%%%%%%%%%%%%%%%%% 
%%%%%%%%%%%%%%%%%%%%%%%%%%%%%%%%%%%%
\section{Introduction} The concept of quantum quench, i.e.,  the sudden change in the Hamiltonian parameters of an  isolated quantum system~\cite{Calabrese_PRL_2006,Polkovnikov_RMP_2011,Eisert_NP_2015,Mitra_ARCMP_2018}, has an extraordinary impact in   both technological applications and fundamental physics.~Not only does it represent a  basic operational tool for quantum 
state manipulations, it also enables one to tailor   material properties~\cite{Basov_NM_2017} and  quantum phases~\cite{cavalleri_2020}.
Furthermore, because a quench drives the system  out of equilibrium,  challenging questions have  intrigued many scientists in the last years: Can the system ``thermalize'' in some sense at long times and, if so, what are the properties of the steady state?   The answers to these non trivial problems  mainly depend~on~two~aspects.   
 First,  the type of quench: While early studies  considered    quenches of   spatially homogeneous parameters~\cite{Cazalilla_PRL_2006,Iucci_PRA_2009,Calabrese_PRL_2011,Mitra_PRL_107,Heyl_PRL_2013,Karrasch_PRL_2012,Kennes_PRB_2013,Collura_PRB_2015,Porta_PRB_2016,Calzona_PRB_2017}, recent works have focused on {\it inhomogeneous} %can the italics be removed?
  quenches  such as  extensive disorder potentials \cite{Santoro_PRB_2013,Santoro_PRL_2012}, e.g., in view of many-body localization~\cite{Tang_PRB_2015,Abanin_AP_2017}, and spatially localized perturbations~\cite{Vasseur_PRL_2013,Schiro_PRL_2014,Kennes_PRB_2014,Kjall_PRL_2014,Weymann_PRB_2015,Bidzhiev_PRB_2017,Ashida_PRL_2018,Bond_PRB_2019,Cavaliere_PRB_2019,goldstein_PRB_2019,Fogarty_PRL_2020}, which can for instance generate  persistent oscillations in physical observables  thus preventing the reaching of a steady state~\cite{Santoro_PRL_2012,Santoro_PRB_2013,Cavaliere_PRB_2019}. 
The second important ingredient in the problem is the type of system. In particular, in the case of integrable quantum systems~\cite{Caux_JSM_2011}, the post-quench dynamics 
 is restricted by a complete set $\{\hat{I}_{\alpha}\}$ of local constants of motions  commuting with the post-quench Hamiltonian~\cite{kinoshita_2006}.  
This implies that, if an out of equilibrium steady state is reached, it can be described by a Generalized Gibbs Ensemble (GGE) density matrix~\cite{Rigol_PRA_2006,Cazalilla_PRL_2006,Cramer_PRL_2008,Wouters_PRL_2014,Vidmar_JSM_2016,Porta_PRB_2018a,Porta_PRB_2018,Ishii_PRE_2019,Porta_SCIREP_2020} 
\begin{equation}\label{rho-GGE}
\hat{\rho}_{GGE}=\frac{\exp (-\sum_\alpha \lambda_\alpha \hat{I}_\alpha )}{{\rm Tr}\left[\exp (-\sum_\alpha \lambda_\alpha \hat{I}_\alpha   )\right]}\, ,
\end{equation}
where the Lagrange multipliers $\{\lambda_{\alpha}\}$  are determined by  the pre-quench state and uniquely characterize the GGE. 

{On the theoretical side, there is a growing consensus that the GGE hypothesis works both for homogeneous~\cite{Rigol_PRA_2006,Cazalilla_PRL_2006,Cramer_PRL_2008,Wouters_PRL_2014,Vidmar_JSM_2016,Porta_PRB_2018a,Porta_PRB_2018,Ishii_PRE_2019,Porta_SCIREP_2020} and inhomogeneous~\cite{Santoro_PRL_2012,Santoro_PRB_2013,Gramsch_PRA_2012,He_PRA_2013,Modak_NJP_2016} quenches. However, only a few experimental GGE signatures have been observed so far, mostly limited to  trapped one-dimensional Bose gases~\cite{Langen_SCI_2015}.  As far as Fermi systems are concerned, the  proposals for GGE detection are based on homogeneous quenches of the interaction strength~\cite{kennes-meden_PRL_2014,sassetti-cavaliere-citro_PRB_2016} and have not found experimental evidence yet. Different 
schemes are thus~needed.   }

A particularly illuminating case where sound results concerning GGE are known is when the post-quench Hamiltonian $\hat{\mathcal{H}}$ is   a quadratic form in the creation and annihilation operators, i.e.,  a one-body operator~\cite{Santoro_PRL_2012,Santoro_PRB_2013,Gluza_SP_2019,Murthy_PRE_2019,yuasa_2019}. In such a case, the latter can always be brought  into a diagonal form   $\hat{\mathcal{H}}=\sum_\alpha \varepsilon_\alpha \hat{\gamma}^{\dagger}_{\alpha} \hat{\gamma}^{}_\alpha$
through a change of basis to   suitable  creation/annihilation operators $\hat{\gamma}^{\dagger}_{\alpha}$,  $\hat{\gamma}^{}_\alpha$ of single particle states $\alpha$ and  the complete set of constants of motion $\{\hat{I}_\alpha\}$ are identified as
 the   number operators  $\hat{n}_\alpha \equiv  \hat{\gamma}^{\dagger}_{\alpha} \hat{\gamma}^{}_\alpha$~\cite{nota-locality}.

The analysis of these systems provides useful insights on fundamental questions. 
In particular, the way  quantum dynamics is   described by a GGE heavily depends on the type of inhomogeneities that are possibly quenched in the system. On the one hand, quenching an extensively dense disorder prevents the system from reaching a strict stationarity, and only long time \textit{time-averages} %can the italics be removed?
 of one-body observables equal the GGE statistical average over Equation~(\ref{rho-GGE})~\cite{Santoro_PRL_2012,Santoro_PRB_2013}. On the other hand, recent results have shown that, if the localized states of~$\hat{\mathcal{H}}$ are sufficiently spatially separated, i.e., if disorder is rare and weak enough, the expectation values  of local observables tend   \textit{in time}   to the ones prescribed by the GGE density matrix~\cite{Murthy_PRE_2019}.

Importantly, in the case of   post-quench one-body Hamiltonians,  one can also {\it quantify} the deviation of GGE from thermal equilibrium.~This can be done  through the single-particle reduced density matrix stemming from $\hat{\rho}_{GGE}$, 
which is explicitly given by $\hat{\rho}_D=\sum_\alpha |\alpha\rangle \langle \alpha| f_\alpha$ and is thus called the ``diagonal ensemble''  in the $\alpha$-basis. Here $f_\alpha\equiv \langle \hat{n}_\alpha \rangle_\circ={\rm Tr}[\hat{n}_\alpha \hat{\rho}_\circ]$  represent the occupancies of the post-quench constants of motion over the pre-quench state $\hat{\rho}_\circ$. They are  in one-to-one correspondence  with the   $\{\lambda_\alpha\}$, which are fixed through the relation $\langle \hat{n}_\alpha \rangle_{GGE}=\langle \hat{n}_\alpha \rangle_\circ$. 
In particular, for fermionic systems, this implies $f_\alpha=(1+\exp[\lambda_\alpha])^{-1}$.
Thus, while the equilibrium state at temperature $T$ and chemical potential $\mu$  corresponds to the Fermi distribution $f^{eq}_\alpha=f^{eq}(\varepsilon_\alpha) =\left\{1+\exp[(\varepsilon_\alpha-\mu)/k_B T]\right\}^{-1}$, or equivalently  to $\lambda^{eq}_\alpha=(\varepsilon_\alpha-\mu)/k_BT$, the out of equilibrium state is characterized by the actual set   $\{f_\alpha\}$,  or equivalently by the set   $\{\lambda_\alpha\}$, and is thus quantified in terms of ``how many'' occupancies $f_\alpha$  deviate from $f^{eq}_\alpha$  and by ``how much''. 

In this paper we focus on quadratic  Fermi systems and address the following question: what is the 
 most elementary  deviation from equilibrium that can produce observable effects? We shall show that  quenching a {\it spatially localized potential} can lead, under suitable circumstances, to 
 an  out of equilibrium 
state that (i) reaches stationarity and (ii) is described by a GGE distribution where only one  parameter $\lambda_\alpha$ deviates from equilibrium,  corresponding to  an only partially occupied bound state lying {\it below} a continuum of fully occupied extended states. Furthermore, we  show that such condition yields a negative absorption spectrum, also known in optoelectronics as the optical gain, thereby paving the way to observe signatures of GGE through optical measurements.

 The paper is organized as follows. After presenting the model in Section \ref{sec-2}, we first focus on the case of a sudden quench of a rectangular quantum well to provide the proof of concept of the effect. In particular, in Section~\ref{sec-3}, we determine the post-quench out of equilibrium distribution, while in Section~\ref{sec-4} we evaluate the related absorption spectrum, displaying the quench-induced negative absorption peak. Then, in Section~\ref{sec-5}, we generalize these results by including realistic effects, namely a finite switching time and a smooth quantum well potential profile. Finally, in Section~\ref{sec-6}, we draw our conclusions. 
 
%%%%%%%%%%%%%%%%%%%%%%%%%%%%%%%%%%%% 
%%%%%%%%%%%%%%%%%%%%%%%%%%%%%%%%%%%% 
%%%%          MODEL AND POST-QUENCH OCCUPANCIES          %% 
%%%%%%%%%%%%%%%%%%%%%%%%%%%%%%%%%%%% 
%%%%%%%%%%%%%%%%%%%%%%%%%%%%%%%%%%%%
\section{Model and Post-Quench Occupancies  for a Sudden Quench} 
\label{sec-2} In order to illustrate the effect, we consider as a pre-quench system a homogeneous one-dimensional gas of free spinless electrons, described by the    Hamiltonian  
%$\hat{\mathcal{H}}^{\rm pre}=\int dx \hat{\Psi}^\dagger(x)  \hat{p}^2  \hat{\Psi}^{}(x)/2m$
$\hat{\mathcal{H}}^{\rm pre}=-\hbar^2 \int dx \,\hat{\Psi}^\dagger(x)   \, \partial_x^2   \hat{\Psi}^{}(x)/2m $, with $\hat{\Psi}$ denoting the electron field operator.
The system is  initially at equilibrium with a reservoir, at a  temperature $T$ and a chemical potential~$\mu$. This entails that the Fourier mode operators $\hat{c}(k)$ diagonalizing the Hamiltonian, $\hat{\mathcal{H}}^{\rm pre}=\int  dk \,\varepsilon(k) \hat{c}^{\dagger}(k) \hat{c}^{}(k)$, are characterized by 
\begin{equation}
\label{exp-val-pre}
\langle \hat{c}^{\dagger}(k)  \hat{c}^{}(k^\prime)\rangle_\circ =\delta(k-k^\prime) f^{eq}(\varepsilon(k)) \quad,
\end{equation} 
{where   $\varepsilon(k)=\hbar^2 k^2/2m$  is the pre-quench spectrum.~Then, the system is disconnected from the reservoir and,   at  the time $t=0$,  a localized attractive potential $V(x)<0$ is switched on near the origin $x=0$, so that the post-quench  Hamiltonian is   $\hat{\mathcal{H}}^{\rm post}=\hat{\mathcal{H}}^{\rm pre} \,+\int dx \hat{\Psi}^\dagger(x) V(x)\hat{\Psi}^{}(x)$.  For the moment, we shall focus on the case of a  sudden  quench, while the effects of a finite switching time will be considered in Section~\ref{sec-5}.  Notably, while  $\hat{\mathcal{H}}^{\rm pre}$ has a purely continuous spectrum, $ \hat{\mathcal{H}}^{\rm post}$ also displays a discrete set of bound states, spatially localized around the origin, and with energies $\varepsilon_n<0$ ($n=0,1,2\ldots$)  lying below the  continuum  branch $\varepsilon>0$. }

%Two arguments make the post-quench dynamics of this isolated system  intriguing. On the one hand,  as the quench potential is local, the system only experiences a negligibly small energy change in the thermodynamic limit. On the other hand, in such a limit, the Anderson orthogonality catastrophe~\cite{Anderson_PRL_1967} ensures that the many-body ground state of the post-quench Hamiltonian is orthogonal to  the pre-quench one, suggesting quite a different behavior. 
The post-quench dynamics of this isolated system is intriguing in view of two opposite expectations. On the one hand, because the quenched potential is local, the energy change experienced by the entire system is   vanishingly small  in the thermodynamic limit, suggesting that even the post-quench distribution should remain a thermal one, just like the pre-quench state. In particular, if the initial state is the pre-quench ground state, one might expect the system to fall into the post-quench ground state, with all the bound states fully occupied. On the other hand, the Anderson orthogonality catastrophe ~\cite{Anderson_PRL_1967,vonDelft_PRB_2012} ensures that, precisely in the thermodynamic limit, the many-body ground states of the pre- and post-quench Hamiltonians are orthogonal, suggesting a different post-quench distribution.  In order to characterize the out of equilibrium dynamics, we first bring the post-quench Hamiltonian, quadratic in  the fermionic fields $\hat{\Psi}$ and $\hat{\Psi}^\dagger$, to its diagonal form $\hat{\mathcal{H}}^{\rm post}= \SumInt_\alpha \varepsilon_\alpha \hat{\gamma}_\alpha^{\dagger} \hat{\gamma}^{}_\alpha$  through a   unitary transformation. Here the  symbol $\SumInt$ is a compact notation indicating   a summation over the discrete spectrum branch  and an  integral over the continuous spectrum branch. This implies, as observed above, that the out of equilibrium dynamics of the system is governed by a GGE, which is characterized by the set of post-quench occupancies $f_\alpha$ of the constants of motion.  

%%%%%%%%%%%%%%%%%%%%%%%%%%%%%%%%%%%
%%%%%%%%%              FIGURE   1            %%%%%%%%%%%%
%%%%%%%%%%%%%%%%%%%%%%%%%%%%%%%%%%%
\begin{figure*}[]
\centering
\includegraphics[width=\textwidth]{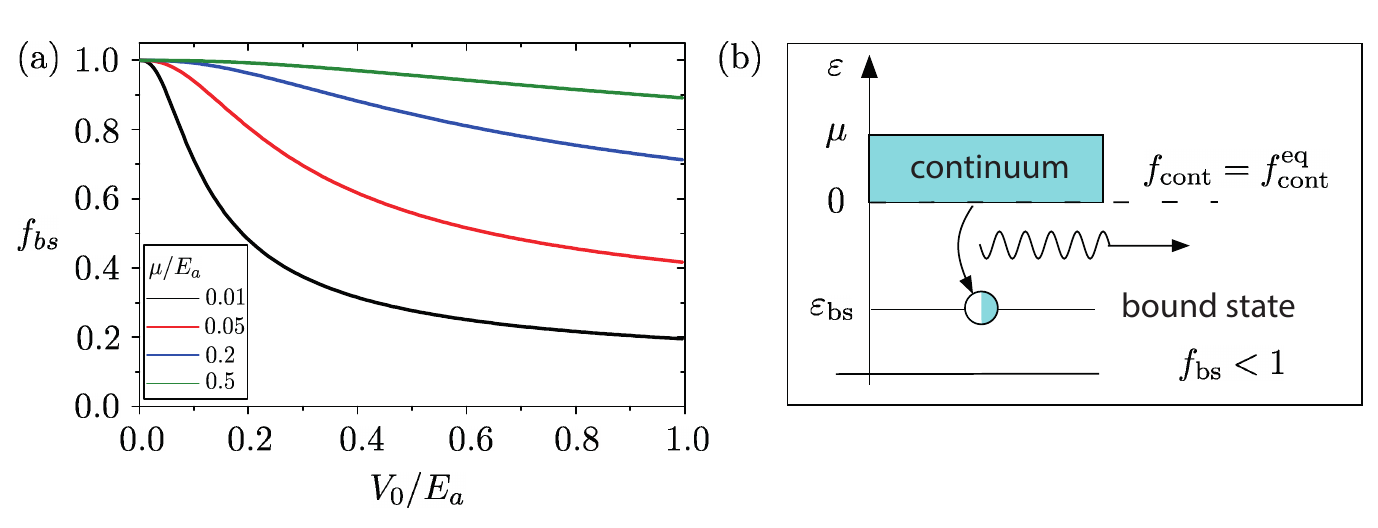}
\caption{(Color online) (\textbf{a}) The occupancy of the bound state $\varepsilon_{\rm bs}<0$ induced by the quench as a function of the quantum well (QW) parameters $V_0/E_a$, at pre-quench  temperature equal to zero, and for four different values of the pre-quench chemical potential $\mu$. (\textbf{b})
Sketch of the occupancy of the post-quench states: While the states of continuum  ($\varepsilon > 0$) are fully occupied up to $\mu$, just like in the pre-quench state, the quench induced bound state gets only partially occupied, realizing the population-inversion regime (optical gain) leading to a stimulated emission of radiation (see Section \ref{sec-4}).  }
 \label{Fig1} 
\end{figure*}
%%%%%%%%%%%%%%%%%%%%%%%%%%%%%%%%%%%

{However, because the post-quench spectrum contains both a discrete and  a continuum branch,  care must be taken in identifying the occupancies $f_\alpha$, which in this case are  determined from the  diagonal ensemble density matrix through the relation $(\hat{\rho}_D)_{\alpha^\prime \alpha}\equiv \langle \hat{\gamma}^{\dagger}_{\alpha} \hat{\gamma}^{}_{\alpha^\prime} \rangle_{GGE} = d_{\alpha \alpha^\prime} f_\alpha$, where $d_{\alpha\alpha^\prime}\equiv \delta_{\alpha\alpha^\prime}$ for $\alpha,\alpha^\prime \in$  discrete spectrum, while $d_{\alpha\alpha^\prime}\equiv \delta(\alpha-\alpha^\prime)$ for $\alpha,\alpha^\prime \in$  continuum  spectrum and $d_{\alpha \alpha^\prime}=0$ otherwise \cite{nota-gamma}. In turn, the $\hat{\rho}_D$ entries can be 
  computed by exploiting the transformation $\hat{\gamma}_\alpha=\int dk \,U(\alpha,k) \, \hat{c}(k)$   linking the post- to the pre-quench operators, where $U(\alpha,k) =\int dx\,\psi^*_\alpha(x) \varphi_k(x)$ is the overlap integral between the post-quench eigenfunctions $\psi_\alpha$ and the pre-quench eigenfunctions $\varphi_k$. 
By recalling the expectation values (\ref{exp-val-pre}) of the pre-quench operators, it is straightforward to show that  }
\begin{equation} \label{occup}
(\hat{\rho}_D)_{\alpha \alpha}=\int dk \left| U(\alpha,k)\right|^2 \, f^{eq}(\varepsilon(k))\quad,
\end{equation}
whence the post-quench occupancies $f_\alpha$ are  obtained through the above prescription. 

%%%%%%%%%%%%%%%%%%%%%%%%%%%%%%%%%%%% 
%%%%%%%%%%%%%%%%%%%%%%%%%%%%%%%%%%%% 
%%%%                   THE CASE OF A QUANTUM WELL                    %% 
%%%%%%%%%%%%%%%%%%%%%%%%%%%%%%%%%%%% 
%%%%%%%%%%%%%%%%%%%%%%%%%%%%%%%%%%%%
\section{The Case of a Quantum Well} 
\label{sec-3}
For definiteness, we shall evaluate the post-quench occupancies for the case of a rectangular quantum well (QW) potential $V(x)=-V_0\, \theta(a/2-|x|)$, characterized by a potential depth $V_0$ and a width $a$ around the origin. Here, $\theta$ denotes the Heaviside function. In this case, space parity is conserved across the quench,  the post-quench   eigenfunctions $\psi_\alpha$ are well known, just like the pre-quench free-particle eigenfunctions $\varphi_k$, and the occupancies Equation~(\ref{occup}) can be evaluated for all the post-quench states. 

As far as the continuous spectrum is concerned, it is worth recalling that  the presence of the QW does modify the continuum  states with respect to the  free-particle waves, especially at  small energies ($0< \varepsilon <V_0$). Nevertheless, a lengthy but straightforward calculation (see Appendix \ref{appendix a} for details), shows that in the thermodynamic limit, the post-quench occupancy of the continuum   is $f_\alpha=f^{eq}(\varepsilon_\alpha)$, i.e.,  it   coincides with the equilibrium Fermi function of the pre-quench state,  with the same temperature and chemical potential,  regardless of the values $a$ and $V_0$ of the QW parameters. This is the hallmark of the locality of the quench. In particular, at zero temperature all continuum states are fully occupied up to the chemical potential $\mu$.

The situation is different for the bound states. As is well known, the number of bound states in a rectangular QW depends on the ratio between the well potential depth $V_0$ and the kinetic energy $E_{a}=\pi^2\hbar^2/2m a^2$ associated to the confinement in the well width $a$. 
The smallest deviation from equilibrium is when one single discrete level, lying below the continuous spectrum of occupied states, is not fully occupied. Moreover, the existence of only one bound state ensures that the system   reaches a stationary state after the quench~\cite{Murthy_PRE_2019}. Focusing then on the regime $V_0<  E_a$, where the QW hosts only one bound state, one can exploit the well known expression for the bound state of a rectangular QW and evaluate its occupancy $f_{\rm bs}=(\hat{\rho}_D)_{\rm bs,bs}$ numerically from Equation~(\ref{occup}). The result is shown in \mbox{Figure~\ref{Fig1}a}, where $f_{\rm bs}$ is plotted as a function of the ratio $V_0/E_a$, at zero temperature, for four  values of chemical potential $\mu$. While for an extremely shallow and thin well ($V_0/E_a \ll 1$) one has $f_{\rm bs} \simeq 1$, i.e., the value one would obtain  if the post-quench system were at equilibrium, 
for $V_0/E_a \lesssim 1$ the occupancy decreases. Notably, such a reduction is, the more pronounced the lower  $\mu$ is, which can be understood from the following arguments. Since the pre-quench eigenfunctions $\varphi_k$ are essentially plane waves,  the  $U({\rm bs},k) $ coefficient is  the Fourier transform  of the bound state wavefunction $\psi_{\rm bs}$ and becomes negligible for $k\gg 1/\ell$, where $\ell \gtrsim a$ is the length scale over which $\psi_{\rm bs}$ is localized. The chemical potential~$\mu$ of the pre-quench state  appearing in the Fermi function, cuts the integral in Equation~(\ref{occup}) at the Fermi wavevector $k_F=\sqrt{2 m\mu}/\hbar$. Thus, while for $k_F\gg 1/\ell$ the occupancy is $f_{\rm bs}=\int dk \left| U({\rm bs},k)\right|^2 f^{eq}(\varepsilon(k))\simeq \int dk \left| U({\rm bs},k)\right|^2 \, =1$ (unitarity of the $U$ transformation), for small chemical potential, such that  $k_F\ll 1/\ell$, the integral is cut before yielding the occupancy  1.

{The resulting occupancy of the post-quench spectrum is  sketched in Figure~\ref{Fig1}b at zero temperature: While the continuum states $\varepsilon > 0$  are characterized by the very same Fermi function as the equilibrium pre-quench state and are thus fully occupied up to the chemical potential~$\mu$ for any QW parameter, the  bound state $\varepsilon_{\rm bs}<0$ is only partially occupied, despite being energetically more favorable than the continuum. This  
 peculiar  out of equilibrium effect  thus realizes the 
 most elementary  GGE deviation from equilibrium: only the bound state $\lambda_{\rm bs}=\ln[(1-f_{\rm bs})/f_{\rm bs}]$ deviates from the equilibrium value.~In particular, this is quite different from the case of a homogeneous quench, where  typically an extensive number of  post-quench occupancies deviate  from equilibrium~\cite{Porta_PRB_2018}. } 

{Note that, because of  particle conservation, the partial occupation of the quench-induced bound state corresponds to an  infinitesimally small depletion (by  at most one electron) of the continuum spectrum. In the thermodynamic limit, no directly seizable effect thus occurs in the  continuum states.
%Note that, because of  particle conservation, the partial occupation of the quench-induced bound state corresponds to a depletion by (at most) one electron of the continuous spectrum, an effect that is not seizable  directly from the continuum itself, since   in the   thermodynamic limit the  number of continuum states is infinite.
 In contrast, the emergence of an only partially occupied bound state, energetically separated from the fully occupied continuum above, has a remarkable consequence: It realizes the condition of population-inversion,   well known   in optoelectronics. While at equilibrium a radiation impinging onto an electron system yields the absorption of an energy quantum causing a transition from  energetically lower and more populated levels to  upper and less populated levels, the out of equilibrium population obtained here leads to a release of energy, causing a stimulated emission or, a ``negative" absorption. This opens up the possibility to observe this GGE signature through optical measurement, as we shall describe in the next section.}

%%%%%%%%%%%%%%%%%%%%%%%%%%%%%%%%%%%% 
%%%%%%%%%%%%%%%%%%%%%%%%%%%%%%%%%%%% 
%%%%                         ABSORPTION SPECTRUM                          %% 
%%%%%%%%%%%%%%%%%%%%%%%%%%%%%%%%%%%% 
%%%%%%%%%%%%%%%%%%%%%%%%%%%%%%%%%%%%
\section{ Absorption Spectrum} 
\label{sec-4}
For an electron system  coupled to an electromagnetic radiation of frequency $\omega$,  the non-linear absorption spectrum $A(\omega)$ is given, within the conventional perturbation-theory based on a Fermi's golden rule treatment of the light-matter interaction~\cite{capasso-book},  by

\begin{eqnarray}\label{alpha-abs}
\lefteqn{A(\omega) 
=
\frac{
 2\pi {\rm e}^2}
{c \,\epsilon_0\,n_\Re   \mathcal{V} m_e^2 \omega}\,\times \hspace{1cm} }& & \\
& & \times \SumInt_{\alpha} \SumInt_{\alpha^\prime} 
\left|\langle \alpha^\prime| p |\alpha \rangle\right|^2
\delta\left(\varepsilon_{\alpha^\prime} -\varepsilon_{\alpha} - \hbar\omega\right)
\left(f_{\alpha} - f_{\alpha^\prime}\right)\nonumber 
\ ,
\end{eqnarray}
%\begin{equation}\label{alpha-abs}
%A(\omega) 
%=
%\frac{
% 2\pi {\rm e}^2}
%{c \,\epsilon_0\,n_\Re   \mathcal{V} m_e^2 \omega} \SumInt_{\alpha} \SumInt_{\alpha^\prime} 
%\left|\langle \alpha^\prime| p |\alpha \rangle\right|^2
%\delta\left(\varepsilon_{\alpha^\prime} -\varepsilon_{\alpha} - \hbar\omega\right)
%\left(f_{\alpha} - f_{\alpha^\prime}\right) 
%\ ,
%\end{equation}
where $p=-i \hbar \partial_x $ is the momentum operator,  $n_\Re$ denotes the real part of the refraction index, $c$ the speed of light, $\epsilon_0$ the vacuum dielectric constant, $m_e$ the bare electron mass and $\mathcal{V}$ the volume.  
Equation~(\ref{alpha-abs}) describes   all transitions from initial states $\alpha$ to final states $\alpha^\prime$  compatible with the transition energy $\hbar \omega$, and its non-linear nature  is determined by the factor $f_{\alpha}- f_{\alpha^\prime}$. While at equilibrium, the final state $\alpha^\prime$ is necessarily less populated than $\alpha$  ($f_{\alpha}>f_{\alpha^\prime} $), causing an actual absorption, $A(\omega)>0$, in the population-inversion regime induced by the quench, one has $f_{\alpha^\prime}>f_{\alpha} $ for 
$\alpha={\rm bs}$ and $\alpha^\prime$ in the occupied continuous spectrum, opening up the possibility of a {\it negative} absorption coefficient, $A(\omega)<0$, i.e., to the emission of an electromagnetic radiation stimulated by the quench.   This is known in optoelectronics as the optical gain effect \cite{capasso-book}. However, unlike the more conventional inter-band transitions, the effect described here can be considered as ``intraband", as it originates from a quench on one single pre-quench band. We also point out that the semiclassical treatment underlying Equation~(\ref{alpha-abs}) is valid for time scales longer than the decoherence time scale, where the density matrix exhibits damped out off-diagonal entries and reduces to the diagonal ensemble. Indeed, as we shall argue in Section~\ref{sec-5}, there exists a finite relaxation time $\tau_{rel}$, after which the density matrix is effectively described by such diagonal ensemble. Thus, within the specified time window, the semiclassical treatment captures the gist of the population-inversion effect.

\subsection*{Implementation}

{As can be deduced from Figure~\ref{Fig1}a, the optimal regime to obtain a population-inversion  is in principle $\mu \ll   V_0 \lesssim E_a$. However, a too small chemical potential reduces screening effects and makes electron--electron interaction effects relevant.~A still quite acceptable regime is  $\mu  \lesssim  V_0 \lesssim E_a$, which  can be achieved, e.g., with an  InSb nanowire (NW), characterized by a small effective mass $m=0.015 m_e$, and   a  realistic  QW    realized by a  finger gate   deposited on a NW portion with size $a=150\,{\rm nm}$ and  biased by a gate voltage $V_0<0$. This yields $E_a \simeq 1.12 \,{\rm meV}$ and, by taking a realistic value $\mu=0.2 \,{\rm meV}$, one still has an energy window for the QW depth $V_0$.~Furthermore, due to the large $g$-factor of InSb NWs \mbox{($g\sim 50$) \cite{kouwenhoven_Nanolett_2013}} the application of a magnetic field of a few Teslas is sufficient to widely spin-split the NW bands, thereby avoiding double occupancy of the bound state, ruling out the related electron--electron interaction effects inside the quantum well.}

\begin{figure}[]

\includegraphics[width=0.5\textwidth]{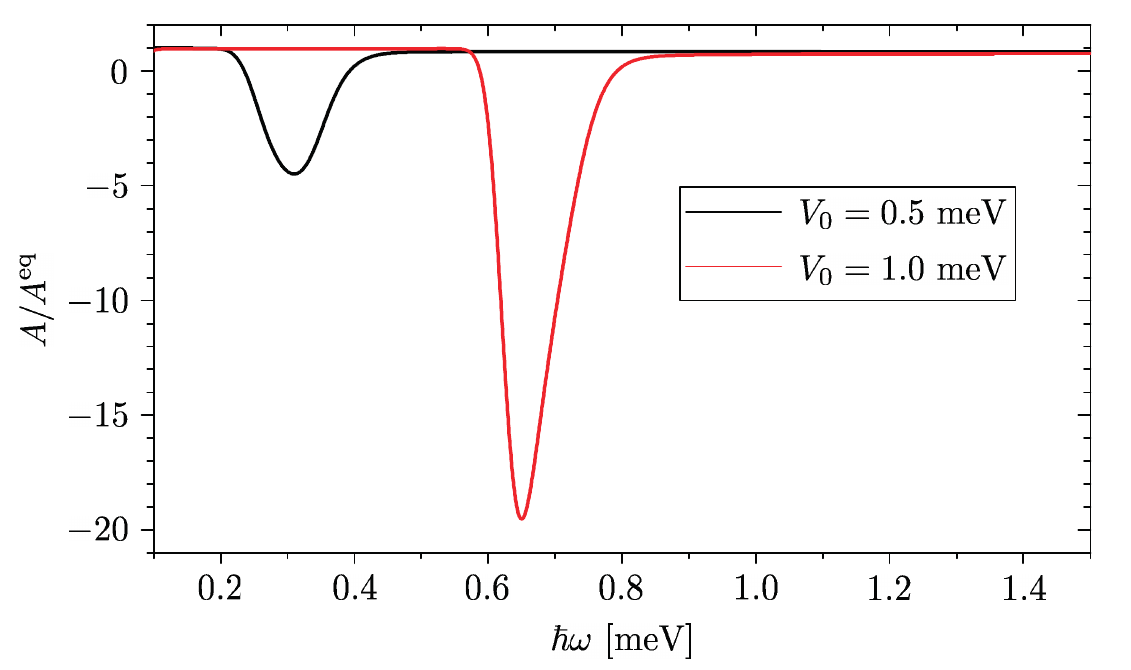}
\caption{(Color online) The ratio $R$ between the out of equilibrium absorption spectrum $A(\omega)$ induced by the quench and the equilibrium  absorption spectrum $A^{eq}(\omega)$ of the post-quench system, for an InSb nanowire (NW) with a QW width $a=150\,{\rm nm}$ ($E_a \simeq 1.12\,{\rm meV}$) and depth $V_0=0.5\,{\rm meV}$ (black curve) and $V_0=1.0\,{\rm meV}$ (red curve). The pre-quench temperature and chemical potential are $T=250{\rm mK}$ and $\mu=0.2\,{\rm meV}$, respectively. While at low frequencies the quench does not induce any deviation from equilibrium ($R\rightarrow 1$), a significant negative peak appears at   $\hbar \omega^*=|\varepsilon_{\rm bs}|$ corresponding to the energy separation between the continuum and the bound state.  }
 \label{Fig2} 
\end{figure}

Since   Equation~(\ref{alpha-abs}) cannot be computed analytically, we have performed a numerically exact evaluation on a finite system, whose total length $L=16\,\mu m$ is  two orders of magnitude bigger than the QW width $a$, at a  realistic  pre-quench  temperature of $T=250\,{\rm mK}$. Furthermore, the unavoidable presence of  inelastic processes   broadening the otherwise sharp energy levels
has been taken into account by replacing the ideal Dirac $\delta$-function appearing in Equation~(\ref{alpha-abs}) with a broadened function of Gaussian shape $\delta(\varepsilon)\rightarrow \delta_b (\varepsilon) = \exp[-\varepsilon^2/2 \varepsilon^2_{\rm b}]/\sqrt{2\pi} \varepsilon_{\rm b}
$, where the value of broadening energy has been taken as  $\varepsilon_{\rm b}=20\,\mu{\rm eV}$. This roughly corresponds to $k_B T$, i.e.,  the typical broadening related to electron--acoustic phonon energy exchange.
The result is illustrated in Figure~\ref{Fig2}, where we have plotted the ratio $R(\hbar\omega)\equiv A(\omega)/A^{eq}(\omega)$ between the out of equilibrium absorption spectrum  induced by the quench and the equilibrium case corresponding to the situation where the post-quench system is at equilibrium, for two different values of QW depth $V_0$.  

%%%%%%%%%%%%%%%%%%%%%%%%%%%%%%%%%%%
%%%%%%%%%              FIGURE   2            %%%%%%%%%%%%
%%%%%%%%%%%%%%%%%%%%%%%%%%%%%%%%%%%
%%%%%%%%%%%%%%%%%%%%%%%%%%%%%%%%%%%

At low frequencies  one has $R(\hbar\omega) \simeq 1$, indicating that the spectrum of the quench-induced absorption coefficient is just like the equilibrium one. In this regime the intraband absorption processes are caused by continuum$\rightarrow$continuum transitions from energetically lower and almost fully occupied states   $0<\varepsilon<\mu$ to energetically higher and almost empty states    $\varepsilon^\prime >\mu$. It is worth pointing out that such transitions occur because of the presence of the QW, which makes the dipole matrix entries $\langle \alpha^\prime| \hat{p} |\alpha \rangle$ non vanishing for $\alpha \neq \alpha^\prime$.

The most interesting effect, however, arises as the frequency approaches the value  $\omega^*\equiv |\varepsilon_{\rm bs}|/\hbar$, where transitions  can occur from the fully occupied lowest continuum states to the only partially occupied bound state lying underneath. This is how  the population-inversion regime  causes a negative absorption, i.e.,  the  stimulated emission of an electromagnetic radiation. The hallmark of this  optical gain effect  is  the  negative peak located around $\hbar\omega^*$. Note that, just like  the value of such resonance frequency, also the depth $R^*$ of the  negative peak is controlled by the value of the potential depth $V_0$, and its magnitude can be significantly higher  than 1, so that the negative absorption is much stronger than the equilibrium positive absorption contribution. For higher frequencies, the ratio $R(\omega)$ becomes positive again. This corresponds to an actual absorption, arising from transitions to the energetically higher and almost empty continuum states from both the bound state and the energetically lower and occupied continuum states.

%%%%%%%%%%%%%%%%%%%%%%%%%%%%%%%%%%% 
%%%%%%%%%%%%%%%%%%%%%%%%%%%%%%%%%%% 
%%%%  SEC. V:  Finite switching time and smooth potential   %%%% 
%%%%%%%%%%%%%%%%%%%%%%%%%%%%%%%%%%% 
%%%%%%%%%%%%%%%%%%%%%%%%%%%%%%%%%%% 
\section{Finite Switching Time and Smooth Potential} 
\label{sec-5}
So far, we have considered the ideal situation of a sudden quench in a quantum well with a sharp rectangular profile.  In realistic implementations, however, the quench is applied over a finite switching time $\tau_{sw}$ and the potential profile of the well is smooth. In this section we thus generalize the results of the previous Sections by taking these aspects  into account. This enables us to demonstrate that the predicted effect relies neither on the instantaneous switching  of the potential nor  on the details of the potential profile, but rather on its property of being local, attractive and hosting a single bound state. Moreover, by simulating the complete time dependent dynamics of the quench, we are able to provide an explicit example of convergence to stationarity of a post-quench local observable, and to show that  its stationary profile is accurately described by the GGE density matrix. 

To this purpose, we now consider a time-dependent  Hamiltonian  $\hat{\mathcal{H}}(t)=\hat{\mathcal{H}}^{\rm pre} \,+g_{sw}(t)\int dx \hat{\Psi}^\dagger(x) V(x)\hat{\Psi}^{}(x)$, where $g_{sw}(t)=\{1+{\rm Erf}[\sqrt{8}\, (t-\tau_{sw})/\tau_{sw}]\}/2$ is a switching function ranging from 0 to 1, up to 2\%, within  a time scale $\tau_{sw}$. Moreover the potential profile $V(x)=-V_0/2 \{ {\rm Erf}[\sqrt{8}\, (x+a/2)/\lambda] - {\rm Erf}[\sqrt{8}\, (x-a/2)/\lambda] \}$ corresponds to a QW with depth $V_0$, width $a$ and edges smoothened over a length $\lambda$. The overall quench dynamics is thus governed by an inhomogeneous and time-dependent Hamiltonian that cannot be treated analytically. By solving numerically the Liouville--von Neumann equation $i\hbar \partial_t \hat{\rho}=\left[ \mathcal{H},\hat{\rho}\right]$ for the single-particle density matrix $\hat{\rho}$, the related diagonal density matrix $\hat{\rho}_D$, associated to the time independent post-quench eigenbasis, is extracted \cite{note-switching}. Differently from the case of a sudden quench, $\hat{\rho}_D$ exhibits a non trivial evolution during the ramp and it becomes constant only after the switching is complete. From $\hat{\rho}_D(t)$ one can then directly observe the time evolution of the occupancies of the post-quench energy levels. 
 
%%%%%%%%%%%%%%%%%%%%%%%%%%%%%
\begin{figure*}[]
\centering
\includegraphics[width=\textwidth]{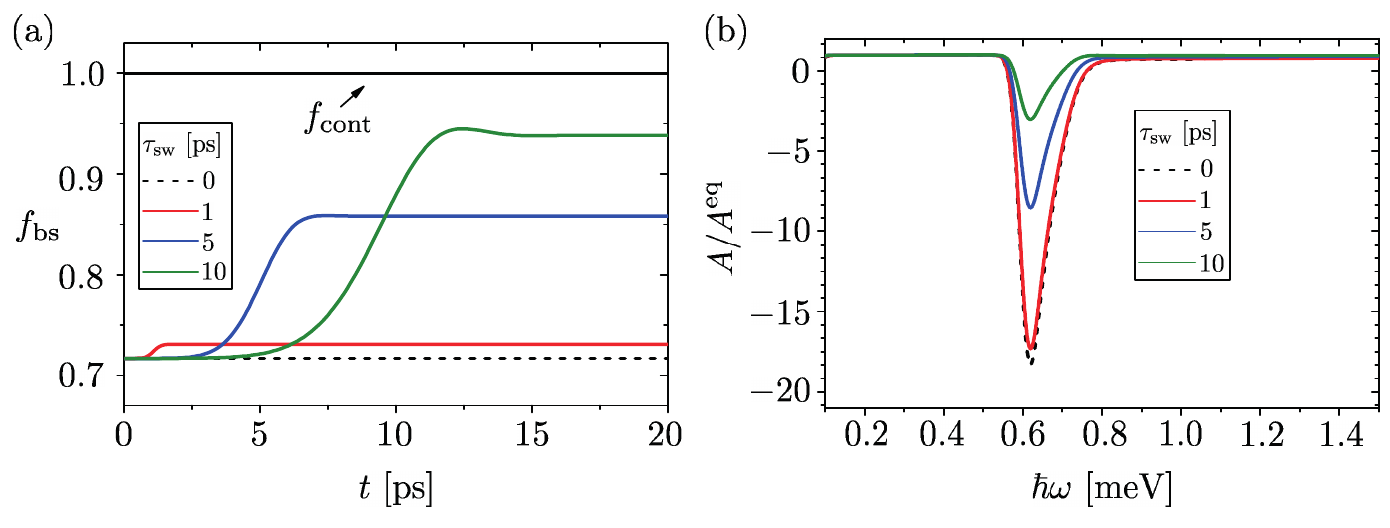}
\caption{(Color online) (\textbf{a}) The time evolution of the occupancy of the bound state, and of the low lying delocalized states, is plotted for  different values of the switching time~$\tau_{sw}$. While the occupancy of the delocalized states remains constant and indistinguishable from $1$  independently on the switching time, the occupancy of the bound state grows from the initial value during the switching time, and reaches a stationary value lower than $1$ after the switching is complete. The dashed curve describes the sudden quench case, for comparison. (\textbf{b}) The ratio $R$ between the out of equilibrium absorption spectrum $A(\omega)$ induced by the quench and the equilibrium  absorption spectrum $A^{eq}(\omega)$ of the post-quench system is shown, at different values of the switching time $\tau_{sw}$. Although the finite switching time reduces the depth of the negative peak with respect to the sudden quench case (dotted curve), its magnitude  remains significantly higher than the  values of the equilibrium spectrum $A^{eq}(\omega)$. In all panels, the computations are performed for an InSb NW with a pre-quench thermal state corresponding to $\mu=0.2\,{\rm meV}$ and $T=250 \,{\rm mK}$ and a post-quench Hamiltonian with a QW potential  of width $a=150\,{\rm nm}$, depth $V_0=1.0\,{\rm meV}$ and a profile smoothening length $\lambda=20\,{\rm nm}$. The energy broadening $\varepsilon_b=20\,\mu{\rm eV}$ has been taken.}
 \label{Fig34} 
\end{figure*}
%%%%%%%%%%%%%%%%%%%%%%%%%%%%%

Taking again as a reference physical system an InSb nanowire of $L=16\,\mu m$ and starting from a thermal pre-quench state with $\mu=0.2\,{\rm meV}$ and $T=250 \,{\rm mK}$, the occupancies of the post-quench bound state and of the post-quench low lying delocalized states (i.e.,  states with energies $ 0< \varepsilon \ll \mu$) are plotted in Figure~\ref{Fig34}a for different values of the switching time~$\tau_{sw}$. The QW parameters are $a=150\,{\rm nm}$ (width ),  $V_0=1.0\,{\rm meV}$ (depth) and $\lambda=20\,{\rm nm}$ (smoothening length). Several features are noteworthy. 
The occupancy of the low lying delocalized states is always indistinguishable from~$1$, independently of the switching time $\tau_{sw}$, as one can see from the thick black horizontal line of Figure~\ref{Fig34}a. In fact, one can verify that the overall distribution of the post-quench delocalized states does not appreciably differ from a thermal one, consistently with the analytical result found for the sudden quench of the rectangular QW in the thermodynamic limit. In contrast, the dynamical behavior of the bound state occupancy   does depend on the finite switching time, as shown by the colored solid curves. In particular, while at $t=0$ it always coincides with the occupancy found in a sudden quench (dashed line), it grows during the ramp and then saturates to a higher value once the switching is complete. Note that the longer the switching time~$\tau_{sw}$, the higher is the final occupancy of the bound state, consistently with the picture that an infinitely slow dynamical evolution favors the  system  relaxation to a lower energy state with the lowest available level being fully occupied. However, for finite but realistic switching time values (see colored curves in Figure~\ref{Fig34}a), the occupancy  of the bound state is still lower than 1 by an appreciable fraction, confirming the above  described picture of a partially occupied bound state lying underneath a continuum of fully occupied delocalized states. 

The robustness of the resulting population inversion effect is supported by the analysis of the absorption spectrum.  Specifically, the `post-quench' absorption spectrum, i.e.,  the value of Equation~(\ref{alpha-abs}) evaluated at time $t\gg \tau_{sw}$ and normalized to the equilibrium absorption spectrum, is shown in Figure~\ref{Fig34}b for various  switching time values. By increasing~$\tau_{sw}$, the shape of the negative peak is roughly unaltered, whereas its depth $R^*$ is reduced. The value $R^*\simeq -18$ obtained for an ideally instantaneous quench (dashed curve in \mbox{Figure~\ref{Fig34}b}) reduces  to $R^*\simeq -17$ (red curve), $R^*\simeq -8$ (blue curve) and $R^*\simeq -3$ (green curve)  for $\tau_{sw}$ values of $1$, $5$ and $10\,{\rm ps}$, respectively. Yet, the value $|R^*|>1$  indicates that the out of equilibrium contribution of the negative peak is still larger than the positive equilibrium one.  This
  clearly visible negative peak  is thus a stable signature of the predicted  out of equilibrium GGE distribution.

Having addressed the robustness of this GGE distribution, we conclude this section by explicitly showing that the unitary dynamics following the quench effectively generates a stationary state whose local properties are well captured by the GGE density matrix. In doing so, we thus provide not only an explicit numerical confirmation of the analytical results found for sudden quenches \cite{Gluza_SP_2019,Murthy_PRE_2019}, but also their generalization  to the more realistic cases with finite switching times. In particular, we have focused on the spatial profile of the charge density.
Note that its  time evolution is characterized by three time scales: the switching $\tau_{sw}$, the recurrence $\tau_{rec}$ and the relaxation time $\tau_{rel}$. The first one depends on the chosen quench protocol and  determines the time after which $\hat{\rho}_D(t)$ becomes stationary. The second one,  associated to the recurrences emerging in any finite size system, scales with the system size and thus tends to infinity in the thermodynamic limit. Finally $\tau_{rel}$ is the time after which the expectation values of any local observable should be reproduced by the stationary $\hat{\rho}_D(t\gg\tau_{sw})$, up to the chosen  accuracy. For sudden quenches it is known that, under suitable hypotheses, a sufficiently large system size always enables one to find a relaxation time $\tau_{rel}<\tau_{rec}$, thus identifying a finite time window in which the state becomes effectively stationary and is accurately described by the GGE density matrix~\cite{Gluza_SP_2019,Murthy_PRE_2019}.

\begin{figure}[]
\includegraphics[width=0.5\textwidth]{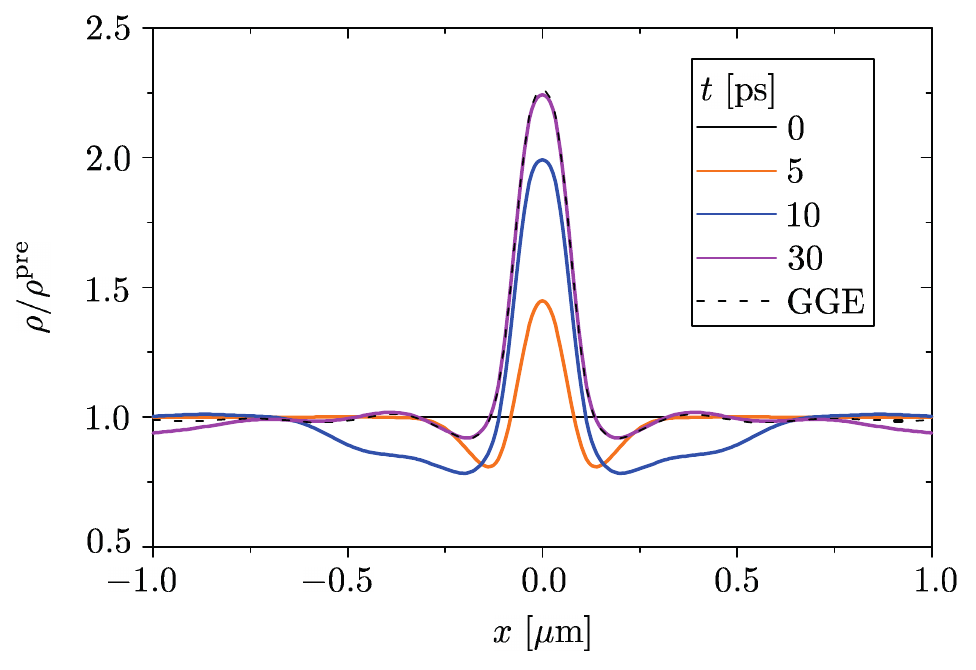}
\caption{(Color online) Different snapshots of the post-quench charge density profile, normalized to the pre-quench spatially uniform profile, are shown for an InSb NW with the same parameters as in Figure~\ref{Fig34}, for a  switching time $\tau_{sw}=5 \,{\rm ps}$ of the QW potential. The charge density profile relaxes to the stationary distribution predicted by the Generalized Gibbs Ensemble (GGE) within a finite time $\tau_{rel}\simeq  30 \,{\rm ps}>\tau_{sw}$. It remains in this out of equilibrium distribution until the recurrence time related to finite size of the system.}
 \label{Fig5} 
\end{figure}

Here we test this prediction in our model for a finite switching time. The charge density profile, renormalized to the pre-quench  spatially uniform distribution, is reported in Figure~\ref{Fig5} for a finite switching time $\tau_{sw}=5\,{\rm ps}$,   at various snapshots. 
During the switching  of the potential ($0< t <2 \tau_{sw}$) the charge density profile starts to deviate  from the pre-quench uniform profile. After the switching has been completed ($t>2\tau_{sw}$), it gradually relaxes  in a finite time to the stationary profile predicted by $\hat{\rho}_D(t\gg\tau_{sw})$. In particular, within a spatial region of 2 $\mu m$ including the QW, the convergence to the GGE occurs within $\simeq$30 $\,{\rm ps}$ (magenta curve in Figure~\ref{Fig5}). The convergence to the GGE result thus holds for a finite switching time as well, with the necessary further restriction $\tau_{sw}<\tau_{rel}<\tau_{rec}$.  We thus conclude that, in the prescribed time window $\tau_{rel}<t<\tau_{rec}$, the state of the system is effectively stationary and indistinguishable from a GGE density matrix. We emphasize that, since the {\it local}    quenched potential is set to induce one single bound state, the convergence to the GGE-prescribed profile occurs {\it in time}, not just upon time-average as in the case of an extensively distributed disorder potential \cite{Santoro_PRB_2013,Santoro_PRL_2012}.
Furthermore, such out of equilibrium GGE distribution corresponds to a very peculiar population-inversion regime, which leads to a unique fingerprint in the absorption spectrum of the system.

\vspace{6pt}

\section{Conclusions} 
\label{sec-6}
We have shown that, by quenching a suitable local attractive potential in an isolated one-dimensional free electron gas, the out of equilibrium dynamics is determined by a GGE  describing the elementary  deviation from equilibrium, where only one Lagrange multiplier $\lambda_{\rm bs}$ deviates from its equilibrium value. 
The  proof of concept of this prediction has been provided in Sections~\ref{sec-3} and \ref{sec-4}, where we have considered the case of a sudden quench in a rectangular QW. This case can be analytically evaluated and the resulting post-quench GGE distribution has been computed exactly.  We have found that the occupancy  of the continuum states is unaltered by the quench and is still described by an equilibrium Fermi function, so that all such states are occupied up to the chemical potential at zero temperature. In striking contrast,   the bound state generated by the quench is only partially occupied, despite being energetically more favorable than the continuum (see \mbox{Figure~\ref{Fig1}}). Such population-inversion regime has been shown to cause a negative peak in the absorption spectrum,  realizing an optical gain  (see Figure~\ref{Fig2}).  The implementation  in   InSb NWs has also been discussed. 

Then, in Section~\ref{sec-5}, we have  considered the more realistic case of a QW potential that is switched on in a finite switching time  and that exhibits  a smooth profile. This analysis, which has been carried out numerically, confirms the robustness of the predicted effect. Indeed, as shown in Fig.~\ref{Fig34}a, the  population of the bound state saturates, after a transient time, to a value that is lower than 1 and that depends on the switching time. The negative absorption thus persists, as shown in Fig.~\ref{Fig34}b. Furthermore, we have also proven in Figure~\ref{Fig5} that, since the {\it locally} quenched potential is set to induce  one single bound state,  the spatial profile of the electron density  tends to the  profile prescribed by the out of equilibrium GGE  {\it in time}, not just upon time average like in the cases of extensively distributed quenched potentials.
In conclusion, these results  based on a  local quench protocol  could  pave the way to observe via optical measurements signatures of GGE in fermionic systems, which have been elusive so far within proposals based on  homogeneous quenches.
%%%%%%%%%%%%%%%%%%%%%%%%%%%%%%%%%%%%%%%%%%
%
%%%%%%%%%%%%%%%%%%%%%%%%%%%%%%%%%%%%%%%%%%
%\authorcontributions{L.R., F.D. and F.R. conceived the project and performed the calculations. All of the authors analyzed and interpreted the results and contributed to writing the paper. All authors have read and agreed to the published version of the manuscript.}
%
%\funding{This research received no external funding.}
%

\acknowledgments{L.R. acknowledges fruitful discussions with T. Monnai. Computational resources were provided by hpc@polito at Politecnico di Torino.}

%\conflictsofinterest{The authors declare no conflict of interest.} 

%%%%%%%%%%%%%%%%%%%%%%%%%%%%%%%%%%%%%%%%%%
%% Optional

%%%%%%%%%%%%%%%%%%%%%%

%%%%%%%%%%%%%%%%%%5

\onecolumngrid
\newpage

\appendix

%%%%%%%%%%%%%%%%%%%%
%%%%%%%%%%%%%%%%%%%5

\section{Post-Quench Occupancies of the Continuum Spectrum for a Suddenly Quenched  Rectangular Quantum Well}
\label{appendix a}
 
%%%%%%%%%%%%%%%%%%%%%
%%%%%%%%%%%%%%%%%%%%%
%%%%%%%%%%%%%%%%%%%%%
\subsection{Continuum Spectrum Eigenfunctions of the Post-Quench Hamiltonian}
As is well known, since the post-quench Hamiltonian 
\begin{equation*}
\hat{\mathcal{H}}^{\rm post}= \int dx \Psi^\dagger(x) \left( -\hbar^2/2m \,   \partial_x^2 \, -\, V_0\, 
\theta(a/2 -|x|)\,\right) \Psi(x)  
\end{equation*}
commutes with the space parity operator, the single-particle eigenfunctions can be classified according to their parity $\eta=\pm=\mbox{even/odd}$. In particular, within a given parity sector $\eta$, the  continuum spectrum ($\varepsilon>0$) wavefunction $\psi_{\eta}(x)$ {\it outside} the QW can be written as a linear combination of two wavefunctions, namely the free-particle wavefunction $\varphi_{\eta}(x)$   and a {\it singular} wavefunction $\bar{\varphi}_{\eta}(x)$, both with the same parity $\eta$. The weight of such linear combination is determined by an angle $\theta_{\eta}$. 
 Explicitly,   denoting by $ q= \sqrt{2 m \varepsilon }/\hbar$ the wavevector outside the QW and by $\tilde q= \sqrt{2 m (\varepsilon+V_0)}/\hbar$ the wavevector inside the  QW, one can label the post-quench unbound eigenfunctions with the discrete-plus-continuum index $\alpha=(\eta,q)$ and compactly write

\begin{eqnarray}
\lefteqn{\psi_{\eta, q}(x)  =} \qquad \qquad 
\left\{ 
\begin{array}{lcl} 
\displaystyle 
\cos\theta_{\eta, q}  \,\, \varphi_{\eta,q}(x) -  \eta \sin\theta_{\eta, q} \,\, \bar \varphi_{\eta ,q}(x) & & |x| \ge a/2 \\ 
 & & \\ 
\displaystyle 
\sqrt{ 
\frac{1+\tan^{2\eta} \big( \frac{\tilde q a}{2} \big) }
{1+ \big(\frac{\tilde q}{q} \big)^2 \tan^{2\eta} \big( \frac{\tilde q a}{2} \big)}
} 
\, \varphi_{\eta,\tilde q}(x)& & |x|<a/2 
\end{array}  
\right.   \label{uno}
\end{eqnarray}
where 
\begin{equation*}
\varphi_{+,q}(x)=\frac{1}{\sqrt{\pi}}\cos(q x)\quad,\quad\varphi_{-,q}(x)=\frac{1}{\sqrt{\pi}}\sin(q x)    
\end{equation*}
are the pre-quench even/odd eigenfunctions, respectively, and 
\begin{equation*}
\bar{\varphi}_{+,q}(x)=\frac{1}{\sqrt{\pi}}\sin(q|x|)\quad,\quad\bar{\varphi}_{-,q}(x)=\frac{1}{\sqrt{\pi}}\mathrm{sgn}(x) \cos(q x)    
\end{equation*}
are their even and odd singular counterparts. Moreover, the angle determining their relative weight in the first line of Equation~(\ref{uno}) is  
\begin{equation*}
\theta_{\eta,q} =  \displaystyle \arctan \left[ \left(\frac{\tilde q}{q}\right)^{\eta}\tan\left(   \frac{\tilde q a}{2}  \right) \right] \,-  \frac{q a}{2}+\, \pi\, \frac{1-\mbox{sgn}\left( \cos\left(\frac{\tilde q a}{2 } \right) \right)}{2} \quad .
\end{equation*}
From the  above definitions, one can then verify that the normalization $\langle   \psi_{\eta,q} | \psi_{\eta', q'} \rangle = \delta_{\eta,\eta'} \delta(q-q')$ holds.
 
%%%%%%%%%%%%%%%%%%%%%%%%%%%%%%%%
%%%   Basis change coefficients (continuum spectrum)    %%%
%%%%%%%%%%%%%%%%%%%%%%%%%%%%%%%%
\subsection{Basis Change Coefficients (Continuum Spectrum)}
In computing the  coefficients of the pre-post basis change appearing in Equation~(\ref{occup}), we observe that $U_{\eta \eta'} (q,k)=\delta_{\eta \eta'}\langle   \psi_{\eta, q} | \varphi_{\eta, k} \rangle$ and
% \begin{widetext}
\begin{eqnarray}
U_{\eta \eta'} (q,k) &=&   
\delta_{\eta \eta'} \Bigg[\underbrace{
\int_{|x|\ge\frac{a}{2}}  \psi^*_{\eta, q} (x) \, \varphi_{\eta, k} (x) \, dx
}_{=C_{\eta}^{out}(q, k)}
\,\,+\,\, 
\underbrace{
\int_{|x|<\frac{a}{2}}  \psi^*_{\eta, q} (x) \, \varphi_{\eta, k} (x) \, dx
}_{=C_{\eta}^{in}(q, k)} \Bigg]  \label{due} \nonumber \\
&\simeq &2\delta_{\eta \eta'} \Bigg[\cos\theta_{\eta,q} 
\left(\int_{\frac{a}{2}}^{\infty}     \varphi_{\eta, q} (x) \, \varphi_{\eta, k} (x)\, dx \right)\nonumber\\ 
&-&\eta \sin\theta_{\eta , q}  
\left(\int_{\frac{a}{2}}^{\infty}  \bar{\varphi}_{\eta, q} (x) \,\varphi_{\eta, k} (x)\, dx  \right) \Bigg]   \label{Cout-gen}
\end{eqnarray}
%\end{widetext}
where in Equation~(\ref{due}) we have neglected the second contribution $C_{\eta}^{in}$, which is negligible with respect to the first contribution $C_{\eta}^{out}$, because the space region outside the QW is infinitely long in the thermodynamic limit and because we are focusing on  the continuum spectrum wavefunctions. Thus, the $U_{\eta \eta'} (q, k)$ coefficients can be straightforwardly evaluated by inserting the definitions of $\varphi$ and $\bar{\varphi}$ into Equation~(\ref{Cout-gen}), and  by exploiting the~identity
\begin{equation*}
\int_{\frac{a}{2}}^\infty e^{(i k-k_{min})x}\, dx = e^{(i k-k_{min})\frac{a}{2}} \left( \frac{k_{min}}{k_{min}^2+k^2} \,+i  \frac{k}{k_{min}^2+k^2}\right)\sim   e^{i  \frac{k a}{2}} \left[\pi \delta(k) +i \,{\rm P.V.}\left(\frac{1}{k}\right) \right]  
\end{equation*}
where \\
\begin{equation}
\left\{
\begin{array}{l}
\displaystyle \delta(k) =  \frac{1}{\pi} \frac{k_{min}}{k_{min}^2+k^2}  \\    \\
\displaystyle  {\rm P.V.}\left(\frac{1}{k}\right) = \frac{k}{k_{min}^2+k^2} 
\end{array}\right.\label{regular} 
\end{equation}
are the regularized versions of the $\delta$-function and the Principal Value (${\rm P.V.}$), respectively, while $k_{min}$ is an infrared cut-off controlling the integral divergences and mimicking the inverse total length of the system ($k_{min} \sim 2/L \rightarrow 0$ in the thermodynamic limit $L\rightarrow \infty$).   
Within a few algebraic steps, one obtains 
\begin{eqnarray}
U_{\eta \eta'} (q, k)&=&\delta_{\eta \eta'} \left\{ \cos\theta_{\eta, q}  \,  \delta(q-k)  \, -  \frac{\eta}{\pi} \sin\left[(q+k) \frac{a}{2} +\theta_{\eta,q}\right]\,{\rm P.V.}\left(\frac{1}{q+k}\right)\right.\nonumber\\ &-&\left.\frac{1}{\pi} \sin\left[(q-k) \frac{a}{2} +\theta_{\eta,q}\right]\,{\rm P.V.}\left(\frac{1}{q-k}\right)\right\}\label{U-ris}
\end{eqnarray}

%%%%%%%%%%%%%%%%%%%%%%%%%%%%%%%%%%%%
%%%%   occupancies of the post-quench states  (continuum)    %%%%
%%%%%%%%%%%%%%%%%%%%%%%%%%%%%%%%%%%%
\subsection{Occupancy of the Continuum Post-Quench Eigenstates}
As explained in Section~\ref{sec-2} (see Equation~(\ref{occup})), the  continuum--continuum diagonal density matrix entries  are given by

\begin{eqnarray}
\rho_{\eta \eta}(q,q) = \langle \hat{\gamma}_{\eta}^\dagger(q)  \hat{\gamma}_{\eta}(q) \rangle     
=\int_{0}^{+\infty} dk \left| U_{\eta \eta} (q, k) \right|^2  \,f^{eq}(\varepsilon(k)) \, \quad.\label{sette}
\end{eqnarray}
Inserting Equation~(\ref{U-ris}) in Equation~(\ref{sette}), their evaluation can be carried out and leads to

\begin{eqnarray}
\rho_{\eta \eta}(q,q) &=&   \delta(0) \Biggl\{ \biggl[   \cos^2\theta_{\eta ,q}-  \frac{\eta}{\pi} \underbrace{  \cos\theta_{\eta,q} \sin\biggl(q a +\theta_{\eta,q} \biggr) }_{\mbox{\tiny bounded} } \underbrace{ {\rm P.V.} \left( \frac{1}{q} \right) \frac{1}{\delta(0)} }_{\rightarrow 0} \biggr] f^{eq}(\varepsilon(q))  \,   
 + \nonumber \\
 & & \hspace{1cm} + \frac{1}{\pi^2} \int_0^\infty dk \,   \underbrace{\sin^2\left[\frac{(q+k) a}{2} +\theta_{\eta,q}\right] }_{\mbox{\tiny bounded}}   \underbrace{ {\rm P.V.}^2\left(\frac{1}{q+k}\right) \frac{1}{\delta(0)} }_{\rightarrow 0} f^{eq}(\varepsilon(k))   + \nonumber \\
& &  \hspace{1cm} +  \frac{1}{\pi^2}  \int_0^\infty dk  \,\underbrace{\sin^2\left[\frac{(q-k) a}{2} +\theta_{\eta,q}\right] }_{\mbox{\tiny bounded}}   \underbrace{ {\rm P.V.}^2\left(\frac{1}{q-k}\right)  \frac{1}{\delta(0)} }_{\rightarrow \pi^2 \delta(q-k)}  f^{eq}(\varepsilon(k))+\nonumber\\
& & \hspace{1cm}   + \frac{2\eta}{\pi^2}  \int_0^\infty dk \underbrace{   \sin\left[\frac{(q+k) a}{2} +\theta_{\eta,q}\right] \sin\left[\frac{(q-k) a}{2} +\theta_{\eta,q}\right] }_{\mbox{\tiny bounded}}\cdot\nonumber\\
&&\hspace{1cm}\underbrace{  {\rm P.V.}\left(\frac{1}{p+k}\right)    {\rm P.V.}\left(\frac{1}{p-k}\right)   \frac{1}{\delta(0)} }_{\rightarrow 0 }  f^{eq}(\varepsilon(k))   \Biggr\} 
 \label{laprima} \\ 
 &=& \delta(0)\biggl\{ \cos^2\theta_{\eta,q} \, f^{eq}(\varepsilon(q))  \,     +  \sin^2\theta_{\eta,q} f^{eq}(\varepsilon(q)) \biggr\} \nonumber \\
  & & \nonumber \\
 &=& \delta(0)  \, f^{eq}(\varepsilon(q))  \nonumber,
\end{eqnarray}
where the regularized $\delta$ and ${\rm P.V.}$ are defined in Equation (\ref{regular}). In particular 
$
\delta(0)=\delta(q=0)= (\pi k_{min})^{-1}  \sim  L/2\pi  \rightarrow \infty
$,
as expected, since  the total number of electrons in the continuum   should scale extensively with the system size. By singling out a $\delta(0)$ pre-factor,  one can see that, apart from the $ \cos^2\theta_{\eta,q}$ contribution in the first line of Equation~(\ref{laprima}), the only   term yielding a finite contribution is the squared Principal Value appearing on the third line, due to the  relation
%\begin{eqnarray*}
%\frac{1}{ \delta(0) }\, {\rm P.V.}^2\left(\frac{1}{ q-k }\right)    \equiv     \pi k_{min}   \, %\frac{(q-k)^2}{\left((q-k)^2+k_{min}^2\right)^2} =   \nonumber \\
% =   \pi   \, \underbrace{\frac{(q-k)^2}{(q-k)^2+k_{min}^2} }_{\rightarrow 1}\, %\underbrace{\frac{k_{min}}{(q-k)^2+k_{min}^2}}_{=\pi \delta(q-k)} \nonumber \\  \rightarrow   \pi^2 %\delta(q-k)\quad.
%\end{eqnarray*}
\begin{eqnarray*}
\frac{1}{ \delta(0) }\, {\rm P.V.}^2\left(\frac{1}{ q-k }\right)    &\equiv&     \pi k_{min}   \, \frac{(q-k)^2}{\left((q-k)^2+k_{min}^2\right)^2} =  \pi   \, \underbrace{\frac{(q-k)^2}{(q-k)^2+k_{min}^2} }_{\rightarrow 1}\, \underbrace{\frac{k_{min}}{(q-k)^2+k_{min}^2}}_{=\pi \delta(q-k)} \nonumber \\  
&\rightarrow&   \pi^2 \delta(q-k)\quad.
\end{eqnarray*}
In conclusion, one obtains that the occupancy of the post-quench continuum states equals the equilibrium one. 


\begin{thebibliography}{99}
\bibitem{note-switching} The post-quench Hamiltonian is defined as the time-independent Hamiltonian once the localized potential is completely switched on, i.e. $\hat{\mathcal{H}}^{\rm post}=\hat{\mathcal{H}}(t\rightarrow \infty)=\hat{\mathcal{H}}^{\rm pre} \,+\int dx \hat{\Psi}^\dagger(x) V(x)\hat{\Psi}^{}(x)$. In practice, due to the form of the ramp function $g_{sw}(t)$, one has $\hat{\mathcal{H}}(t \gtrsim2\tau_{sw}) \simeq  \hat{\mathcal{H}}^{\rm post}$.
\end{thebibliography}
\end{document}